\documentclass{aa}
\usepackage{graphicx}
\def\hi{H\,{\sc i} }

\def\kmss{km~s$^{-1}$ }
\def\kms{km~s$^{-1}$}

\def\smppc2{${\rm M}_{\sun} {\rm pc}^{-2}$}

\def\kprime{K^{\prime}}

\begin{document}
\title{MOND rotation curves for spiral galaxies with Cepheid-based
distances}
\author{Roelof Bottema \inst{1}
\and Jose L. G. Pesta\~na
\inst{2}
\and Barry Rothberg
\inst{3}
\and Robert H. Sanders
\inst{1}}
\institute{Kapteyn Astronomical Institute, P.O. Box 800, 
NL-9700 AV Groningen, The Netherlands, e-mail:sanders@astro.rug.nl
\and Dpto. de F\'{\i}sica, Univ. de Ja\'en, Virgen de la Cabeza 2,
   23071-Ja\'en, Spain
\and Institute for Astronomy, 2680 Woodlawn Dr., Honolulu, HI 96822, U.S.A.}
\date{Received 18 February 2002 / Accepted 9 July 2002}
\authorrunning{Bottema et al.}
\titlerunning{MOND rotation curves}
\abstract{
Rotation curves for four spiral galaxies with recently
determined Cepheid-based distances are reconsidered
in terms of modified Newtonian dynamics (MOND).  For two of the objects, 
NGC 2403 and NGC 7331, the rotation curves predicted by MOND
are compatible with the observed curves when these galaxies are
taken to be at the Cepheid distance.
For NGC 3198, the largest distance for which reasonable 
agreement is obtained is 10\% smaller than the Cepheid-based distance;
i.e., MOND clearly prefers a smaller distance.  This conclusion is
unaltered when new near-infrared photometry of NGC 3198 is taken as the
tracer of the stellar mass distribution.  For the large Sc spiral, NGC 2841,  
MOND requires a distance which is at least 20\% larger than the Cepheid-based
distance.  However, the discrepancy of the Tully-Fisher and
SNIa distances with the Cepheid determination casts some doubt upon the
Cepheid method in this case.
\keywords{galaxies: individual: NGC 3198, NGC 2841 --
galaxies: kinematics and dynamics --
galaxies: spiral}
}
\maketitle
\section{Introduction}

It is well established that, in the context of Newtonian dynamics, the 
observable mass in spiral galaxies cannot account
for the observed flat rotation curves in the
outer regions of galaxies (Bosma 1978; Begeman 1987, 
van Albada et al. 1985).  The standard explanation for this discrepancy is
the proposal that galaxies are embedded in
an extended dark halo which
dominates the gravitational field in the
outer regions (Trimble 1987).

An alternative explanation for the discrepancy
is the possibility that dynamics becomes non-Newtonian
in the limit of low accelerations. 
The most successful such proposal is Milgrom's (1983)
modified Newtonian dynamics or MOND.
Here the idea is that below a certain acceleration
threshold ($a_o$) the effective gravitational acceleration
approaches $\sqrt{a_og_n}$ where $g_n$ is the usual
Newtonian acceleration.  This modification yields
asymptotically flat rotation curves of spiral
galaxies and a luminosity -- rotation velocity
relationship of the observed form, $L \propto v^4$,
the Tully-Fisher relation (Tully \& Fisher 1977). But apart from
these general aspects the prescription also successfully
predicts the observed form of galaxy rotation curves
from the observed distribution of stars and gas
with reasonable values for the mass-to-light ratio
of the stellar component (Begeman et al. 1991;
Sanders 1996; Sanders \& Verheijen 1998, McGaugh \& de Blok 1998).
A crucial element of a very
specific prescription like MOND is that the
precise form of the rotation curve is predicted
by the observed mass distribution given the
value of a single universal parameter; in this
case, the critical acceleration $a_0$. Consequently
MOND can in principle be falsified as soon
as there is one galaxy for which the predicted rotation curve
disagrees significantly with the observed curve; although, 
in practice, the usual uncertainties inherent in astronomical data 
render a definitive falsification problematic in any
individual case. 

In Begeman et al. (1991, hereafter
BBS) MOND is applied to
a sample of galaxies for which high
quality \hi rotation curves are available.
For a value of $a_0$ equal to 1.21~10$^{-8}$ cm~s$^{-2}$
the rotation curves of the sample could 
be reasonably reproduced, the free parameter in each case being the
mass-to-light ratio of the visible disc.  Because MOND is an acceleration
dependent modification, this derived value of $a_o$
depends upon assumed distance scale ($H_o = 75$ \kmss Mpc$^{-1}$ in
this case).  Moreover, the quality of an individual fit depends
upon the adopted distance to the galaxy, and, since the relative distances
to these nearby galaxies have not been known to within
an accuracy, typically, of 25\%, this has provided 
some freedom to adjust the distance in order to improve the MOND fit; i.e., 
distance, within certain limits, can be considered as an 
additional second parameter in the fitting procedure.  For most of
the galaxies in the sample of BBS, the distance did not have to
be adjusted significantly ($<10\%$) to improve the MOND fits, and the
improvement was not significant.  However, one object, NGC 2841, required
a large readjustment:  The Hubble law distance to this galaxy is
about 9 Mpc, but MOND clearly prefers a distance which is twice as
large.  

Using ground-based and Hubble Space Telescope observations 
Cepheid distances to 21 inclined galaxies have now been
determined as part of the HST key program on the extragalactic
distance scale (e.g. Sakai et al. 2000).
Three of the galaxies in this Cepheid sample
are also in the sample with high quality
rotation curves considered by BBS. 
These are NGC 2403 (Freedman \& Madore 1988), NGC 3198 (Kelson et al. 1999) 
and NGC 7331 (Hughes et al. 1998).  For
these three galaxies
the MOND prescription can now be considered 
in the context of the Cepheid distance that is generally
considered to be the most precise indicator. 

NGC 2841 has been discussed as a critical case for MOND by Sanders (1996).
For this galaxy, there is also a large discrepancy between the 
Tully-Fisher distance
and the Hubble law distance (for plausible values of the Hubble constant). 
Moreover, the galaxy was the site of a recent SNIa (1999by).  
For these reasons this galaxy has 
been included, subsequently, in the HST program (Macri et al. 2001). 

Here we demonstrate
that for two galaxies in the BBS sample the rotation curve predicted by 
MOND is consistent with the observed curve when the galaxies
are placed at the Cepheid distance.  However,
for NGC 3198 at the Cepheid distance of 13.8 $\pm$ 0.6 Mpc, 
the shape of the rotation curve predicted by MOND 
systematically deviates (by up to 10 \kms) from the observed curve,
both in the inner and outer regions.  The largest distance which can
be compatible with MOND is about 10\% lower than the Cepheid-based distance.
This is not particularly problematic because of likely uncertainties
in the Cepheid method and in the determination of a rotation curve
from the observed two-dimensional velocity field.  
NGC 2841, however, remains a difficult case for MOND.  The minimum
distance which is consistent with MOND is about 17 Mpc whereas
the Cepheid-based distance is 14.1 $\pm$ 1.5 Mpc.  We discuss the 
implications and seriousness of this discrepancy for MOND, or, alternatively,
for the Cepheid method.

\section{Determination of the MOND rotation curve}

The procedure followed when determining a MOND rotation
curve has been described previously (e.g., BBS).
In the context of MOND, the true
gravitational acceleration $g$, is related to the Newtonian
acceleration $g_{\rm n}$ as

\begin{equation}
\mu (g/a_0)g = g_{\rm n}
\end{equation}
where $a_0$ is the acceleration parameter and $\mu(x)$ is
some function which is not specified but has the asymptotic
behavior

\begin{equation}
\mu(x) = 1,\;\;\;\; x > 1\;\;\; {\rm and}\;\;\; \mu(x) = x,\;\;\; x < 1
\end{equation}
(Milgrom 1983);  a convenient function with this asymptotic 
behavior is

\begin{equation}
\mu(x) = x(1 + x^2)^{-1/2}.
\end{equation}
The circular velocity is given as usual by

\begin{equation}
v = \sqrt{ r g}.
\end{equation}
From Eqx. (1), (2) and (4) it is evident that the rotation
curve about a finite bounded mass $M$ in the low acceleration
limit is asymptotically flat at a value given by

\begin{equation}
v^4 = GMa_0,
\end{equation}
which forms the basis of the observed Tully-Fisher (TF) relation.
The Newtonian acceleration
$g_{\rm n}$ is determined, as usual, by applying the Poisson equation
to the mass distribution deduced from the
distribution of the observable matter (disc, bulge, and gas). 
The surface density distribution of the stellar disc is assumed to
be traced by the distribution of visible light (i.e., no variation of
M/L within a given component of a given galaxy), but then the question 
arises as to which
photometric band is most appropriate.  The near-infrared emission
(e.g., $\kprime$-band) is considered to be a better tracer
of the old dominant stellar population, and less susceptible
to position-dependent extinction, but this is not
generally available.  Below, we use the r-band as a tracer
of the form of the mass distribution in the stellar disc, but, with respect
to NGC 3198, we also consider more recent $\kprime$-band photometry.

The stellar disc may be assumed to be asymptotically thin or
have a finite thickness related to the radial scale length of
the disc by an empirical rule (van der Kruit \& Searle 1981);
this makes little difference in the final result.  Applying
Eqx. (1), (3), and (4), a least squares fit is then made to the 
observed rotation curve $v(r)$ where the single free parameter
of the fit is the mass-to-light ratio of the disc; in cases where
there is an indication of a bulge from the light distribution, 
M/L of the bulge enters as a second parameter.

For the gaseous component a surface density
distribution equal to that of the \hi
is taken, multiplied by a factor 1.3 to account for primordial helium.
The gas layer is taken to be infinitesimally thin. The contribution of
the gas to the total rotation is fixed, but does depend on the distance
to the galaxy.

In principle, the parameter $a_o$ should be universal and, having
determined its magnitude, one is not allowed to adopt this as a free 
parameter.  But as noted above, the derived value of $a_o$ does 
depend upon the assumed distance scale.  Sanders and Verheijen (1998) 
give MOND fits to the rotation curves of 30 spiral galaxies in the
UMa cluster which they assume to be at 15.5 Mpc.  The preferred
value of $a_o$ with this adopted distance is equal to the BBS value
of $1.2\times 10^{-8}$ cm~s$^{-2}$.  However, based upon the Cepheid-based
re-calibrated Tully-Fisher relation (Sakai et al. 2000), Tully \& Pierce
(2000) argue that the 
distance to UMa should be taken to be 18.6 Mpc.  
We have recalibrated the Tully-Fisher law using this same sample of galaxies
but with the three test galaxies (NGC 2403, NGC 3198, NGC 7331) left out 
of the fitting.  Within the errors, the slope and intercept of the 
Tully-Fisher relation are the same as that found by Sakai et al.\ (2000),
and the distance to Ursa Major is only 1\% smaller than that found 
Tully \& Pierce (2000).  In that case the MOND fits to the UMa galaxies imply
that the value of $a_o$
should be adjusted to $0.9 \times 10^{-8}$ cm~s$^{-2}$.  This is also the 
preferred value of $a_o$ from MOND fits to rotation curves of a sample of 
nearby dwarf galaxies with distances taken primarily from group membership 
(Swaters \& Sanders 2002, in preparation).

\section{Rotation curve fits at Cepheid-based distances}

In Fig. 1 we show the MOND rotation curve
for all four galaxies from the BBS sample with Cepheid-based distance
determinations.
NGC 2841 and NGC 7331 both contain central bulges as evidenced
in the light distribution, and the radial surface brightness
profile has been appropriately decomposed.  Here, $a_o$ is fixed at the
rescaled value of $0.9\times 10^{-8}$ cm~s$^{-2}$, and the distance is fixed
at the Cepheid-based values as updated and corrected by Freedman et al.
(2001).  The free parameters
of the fit are the disc and, in two cases, bulge masses.  
The resulting values and  
the corresponding mass-to-light ratios are given in Table 1 for the 
four galaxies.  

\begin{figure}
\resizebox{\hsize}{!}{\includegraphics{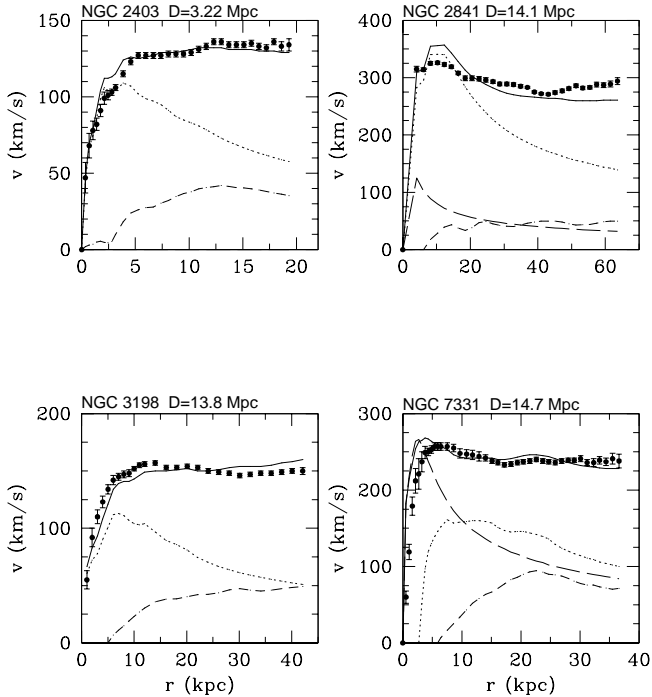}}
\caption[]{
MOND rotation curves compared to observed \hi rotation curves for the
four galaxies from the sample of BBS with Cepheid-based distances.
The dotted, long-dashed, and short-dashed lines are the 
Newtonian rotation curves of the
stellar disc, bulge, and gaseous components respectively.}
\end{figure}
 
Here we see that for two of the galaxies, NGC 2403 and NGC 7331, the
MOND rotation curves agree well with the observed curve.  In both of
these cases, the predicted rotation curve lies somewhat above the
observed curve in the inner regions but this could be due to
beam-smearing.  The implied mass-to-light ratios generally fall within
the range that would be considered reasonable for stellar populations
(Table 1).

\begin{table*}
\begin{flushleft}
\caption{Galaxies with well-defined rotation curves and Cepheid-based  
distances \label{t1}}
\begin{tabular}{|c|c|c|c|c|c|c|c|}
\hline
${\rm Galaxy }$ & ${\rm D}$ & ${\rm L_B}$ 
    & ${\rm M_{gas}}$ & ${\rm M_{disc}}$ & ${\rm M_{disc}/L_B}$ 
    & $ {\rm M_{bulge}}$ & $ {\rm M_{bulge}/L_B}$ \\
$   $ & $({\rm Mpc}$) & $({\rm 10^{10} L_\odot})$ & $({\rm 10^{10} M_\odot})$ 
    & $({\rm 10^{10} M_\odot})$ & $  $ & 
    $ ({\rm 10^{10} M_\odot}) $ & $ $ \\
 (1) & (2) & (3) & (4) & (5) & (6) & (7) & (8) \\ 
\hline
  NGC 2403 & 3.2$\pm$ 0.2 & 0.82 & 0.4 & 1.34 $\pm$ 0.03& 1.6 &  &   \\
  NGC 2841 & 14.1$\pm$ 1.5 & 4.60 & 2.7  & 29.70 $\pm$ 4.3 & 8.3 & 1.5 & 0.83 \\
  NGC 3198 & 13.8$\pm$ 0.5 & 2.44 & 1.6 & 2.63 $\pm$ 0.1 & 1.1 & &  \\
  NGC 7331 & 14.7 $\pm$ 0.6& 5.26 & 1.4 & 13.20$\pm$ 0.6 & 2.0 & 5.7 & 1.8 \\
\hline
\noalign{\smallskip}
\end{tabular}
\\

(2) The Cepheid-based distance from Freedman et al. 2001. \\
(3) The B-band luminosity (in $ {\rm 10^{10} L_\odot}$) 
  at the Cepheid distance. \\
(4) The total gas mass including primordial helium 
  at the Cepheid distance. \\
(5) The total mass of the stellar disc from the 
  MOND fit. \\
(6) The implied mass-to-light ratio of the stellar 
  disc. \\
(7) The total mass of the stellar bulge in those two cases where a bulge
  is evident. \\ 
(8) The implied mass-to-light ratio of the stellar bulge. \\
\end{flushleft}
\end{table*}

For the other two galaxies, there are clear 
systematic differences between the MOND rotation curve and the
observed curves.  Basically, the predicted curves have a different
shape than the observed curves:  
for NGC 2841, the predicted curve is significantly higher
than observed in the inner regions (by up to 30 \kms) and comparably lower 
in the outer regions.  For NGC 3198 the differences are in the opposite
sense:  about 10 \kmss lower in the inner regions and 10 \kmss higher in
the outer regions.  These differences diminish if NGC 2841 is moved further
out and if NGC 3198 is moved closer in; i.e., MOND clearly prefers a
larger distance to NGC 2841 (as discussed previously by BBS and by Sanders
1996) and a smaller distance to NGC 3198.  We now discuss these two cases 
with respect to the question of whether or not this mismatch can
be interpreted as a falsification of MOND.  Because the rotation curve
of NGC 3198, when taken at the Hubble law distance of 10 Mpc, is 
very well predicted by MOND, and because the observed curve is thought
to be well-determined, this, at first sight, appears to be
the more problematic case, and we begin with this object.

\section{NGC 3198}

\subsection{r-band photometry}

This gas-rich spiral galaxy has a generally symmetric \hi
distribution, and there are no large scale significant warps or distortions 
of the velocity field.  The rotation curve extends to roughly 10 radial
scale lengths and is, to first order, flat and featureless (Begeman 1987).
For these reasons it has become the classic case of a spiral galaxy 
evidencing a large mass discrepancy in its outer regions (van Albada
et al. 1985).  If any theory, such as MOND, fails to predict the rotation
curve of this galaxy, then it would be problematic for that theory.

\begin{figure}
\resizebox{\hsize}{!}{\includegraphics{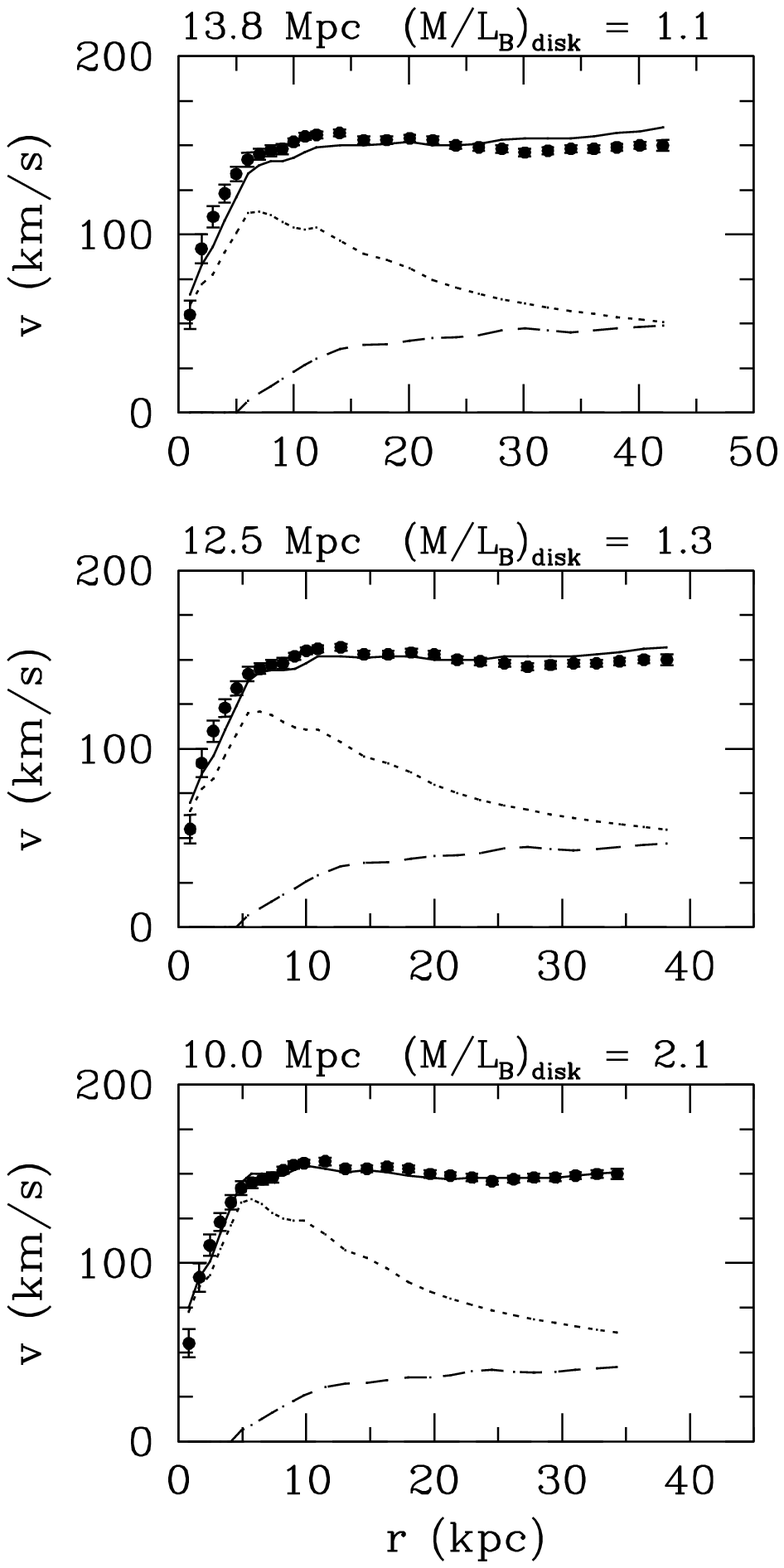}}
\caption[]{
MOND rotation curves for NGC 3198 assumed to be at various distances.
The Cepheid-based distance is 13.8 Mpc.  The dotted and dashed lines are the
Newtonian rotation curves of the stellar and gas discs respectively.
}
\end{figure}

In Fig.\ 2 we show the MOND rotation curves of
of NGC 3198 when the galaxy is assumed to be at distances of
10 Mpc, 12.5 Mpc and 13.8 Mpc.  Again, the MOND acceleration
parameter is assumed to be the BBS value rescaled to the new distance
scale, i.e., $0.9\times 10^{-8}$ cm~s$^{-2}$.

The closest assumed distance, 10 Mpc, is roughly the Hubble law distance
given the radial velocity of NGC 3198 with respect to the local
group;  it is also the least-square-fit distance if distance is
left as a free parameter in the context of MOND.
The distance of 13.8 Mpc is the final Cepheid-based distance
given by Freedman et al. (2001); and 12.5 Mpc corresponds to the
Cepheid distance less 10\%.  The disc M/L values in
the B-band corresponding to the MOND fits at these various distances
are also given in Fig.\ 2.

Here we see that the MOND
rotation curve for a distance of 10 Mpc is essentially a perfect fit 
to the observed 
curve.  At the distance of 12.5 Mpc, the MOND curve is less than a perfect
match, but, nowhere that the rotation
curve is well measured, does the predicted rotation curve deviate by more 
than 5 \kmss from the curve derived from the observed velocity field.  
This is typically
within the difference in the rotation curves derived from the two sides
of the galaxies considered separately-- a sensible estimate of the 
uncertainties (the error bars are formal errors determined from the
tilted ring fitting procedure). 

At the Cepheid distance of 13.8 Mpc, the MOND rotation curve deviates
in the same sense but by now up
to -10 \kmss in the inner regions (8-14 kpc)
and by +10 \kmss in the outer regions (30-40 kpc).  The reason for the
deteriorating fit with increasing assumed distance is the relatively
larger contribution of the gaseous component to the rotational velocity.
The rotation curve of NGC 3198 in the outer regions ($r>20$ kpc) 
is constant at about 150 \kms.  This would imply, in the context of 
MOND, that essentially the entire mass of the galaxy is enclosed within
about 20 kpc, but this is obviously not the case given the significant
surface density of neutral gas in the outer regions-- contributing more
than 50 \kmss to the Newtonian rotation curve at the last measured point.

At a distance of 12.5 Mpc, the MOND rotation curve appears to be consistent
with the observed curve (within the likely errors of the method for estimating
rotation curves from 21 cm line data).  Although this distance is 
formally 2$\sigma$
below the Cepheid-based distance, it is unclear if all systematics
effects connected with this method are well-understood.
It has been noted, for example, that for the galaxy NGC 4258  
the kinematic water-maser-based distance is also about 10\% less than
the Cepheid-based distance (Maoz et al. 1999).  
The error budget of the
Cepheid method is probably on the order of 10\%.

\begin{figure}
\resizebox{\hsize}{!}{\includegraphics{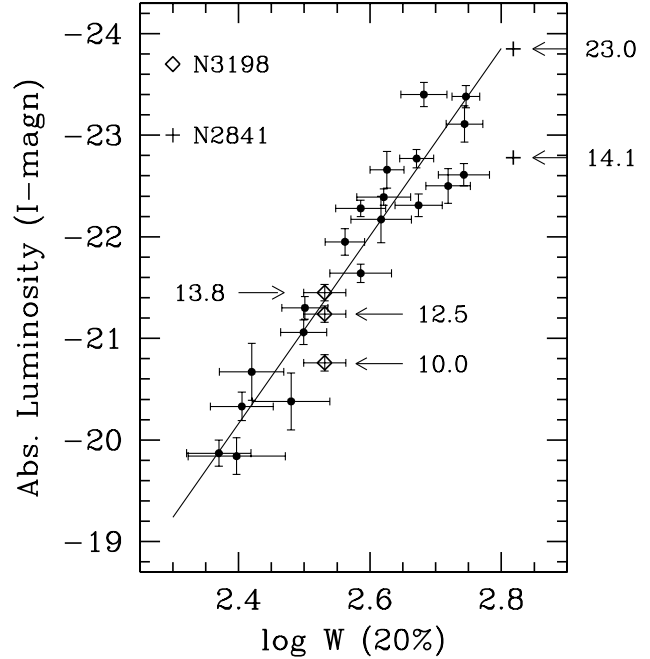}}
\caption[]{
The I-band Tully-Fisher relation for local calibrator galaxies with
Cepheid-based distances.  The positions of NGC 3198 (diamonds)
and NGC 2841 (crosses), 
assumed to be at various distances, are indicated
}
\end{figure}

Sakai et al. (1999) have calibrated the
T-F relation using  21 spiral galaxies with known Cepheid distances 
in five color bands: B, V, R, I, and H.
If one places NGC 3198 on the mean B-band relation
its distance should be 12.2 Mpc, while for the I-band this distance
is 13.3 Mpc. Thus the Tully-Fisher distance is essentially consistent
with the maximum MOND distance.  Although the MOND rotation curve fit
clearly prefers a somewhat
smaller distance than the Cepheid-based distance, the idea is in 
no sense falsified by this well-determined rotation curve.

The I-band Tully-Fisher relation from Sakai et al. is shown in Fig.\ 3.
The open points show the position of NGC 3198 when at a distance of
10.0 Mpc, 12.5 Mpc, and 13.8 Mpc.  It is evident that, given the scatter
in the observed relation, it is impossible to distinguish between these
possibilities although distances of 12.5 to 13.8 Mpc are clearly
preferred.

\subsection{K$^{\prime}$ band photometry}

One possible reason for the small deviation of the MOND curve
from the observed curve at the Cepheid-based distance is that
the r-band photometry is not a precise tracer of the stellar light
distribution due to possible contamination by newly-formed stars
and dust absorption.  For this reason we have also considered
recent near-infrared photometry of this galaxy.

\begin{figure}
\resizebox{\hsize}{!}{\includegraphics[angle=-90]{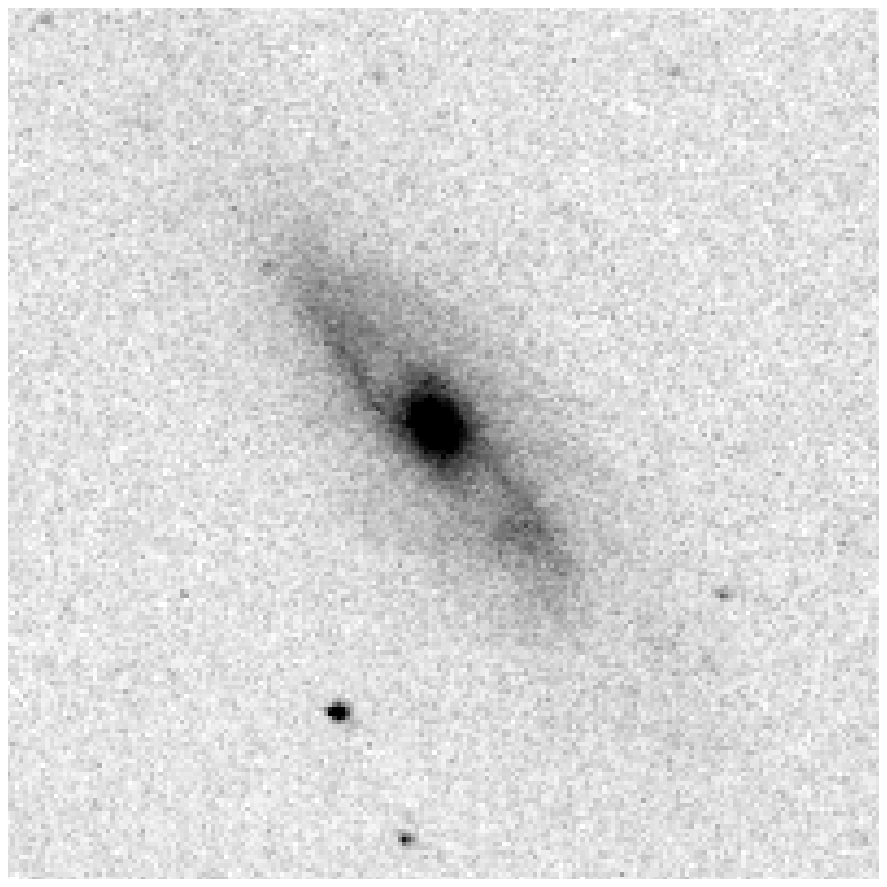}}
\caption[]{
The $\kprime$ image of the central regions of NGC 3198.  Brightness
is represented by a linear gray-scale until 16.94 mag. arcsec$^{-2}$ and
black beyond this level.  The size of the image is 5.6 by 5.6 arc-minutes.
North is at the top, east to the left.  NGC 3198 has a clear bulge in the 
near-infrared surrounded by what seems to be a small light 
depression and then a ring of spiral arm features.
}
\end{figure}

An image of NGC 3198 in the $\kprime$ band has been obtained by
Rothberg et al. (2000) in order to calibrate the near infrared 
Tully-Fisher relation.
The observations and initial stages of the data reduction, like
sky-subtraction and flat-fielding are described in that paper.
The detector was 1024 $\times$ 1024 square pixels
of size 1\farcs{68} $\times$ 1\farcs{68}. Consequently the total
image measures 28.7 arc-minutes along the sides and NGC 3198 which
has a scale-length of approximately 1 arc-minute fits completely within
the image leaving ample margins of pure sky around the galaxy.
Rothberg et al. (2000) derived a total brightness of 7.79 $\kprime$ magnitudes
which translates to 3.4~10$^{10}$ $L_{\sun}^{\kprime}$ for a distance
of 13.8 Mpc.

In Fig. 4 the image of the central regions of NGC 3198
is reproduced. Clearly discernible is a prominent bulge which is much
less obvious in images at bluer wavelengths. As a consequence this
central bulge region must be enshrouded in an appreciable amount of
dust, which explains the reddening going inward. Surrounding the
bulge appears to be a ring of spiral arm features with a light
depression between the bulge and this ring. 

To determine the radial luminosity profile, ellipses have been fitted
to the image which provided the position and orientation of the galaxy
(Fig.\ 5).  As a next step the intensities have been averaged over 
elliptic annuli.
In the inner regions the orientations of the annuli were equal to 
those determined by the ellipse fit, while for the intermediate and
outer regions a constant position angle and ellipticity was adopted.
The error of each radial intensity value was calculated by quadratically
adding the error generated by sky-level variations and the noise
appropriate for each annulus.  The result is shown in Fig.\ 6.
The radial profile
in the r-band is also plotted in that figure, and one may notice that
the photometry of the disc is of similar shape for the r and $\kprime$ bands.

\begin{figure}
\resizebox{\hsize}{!}{\includegraphics{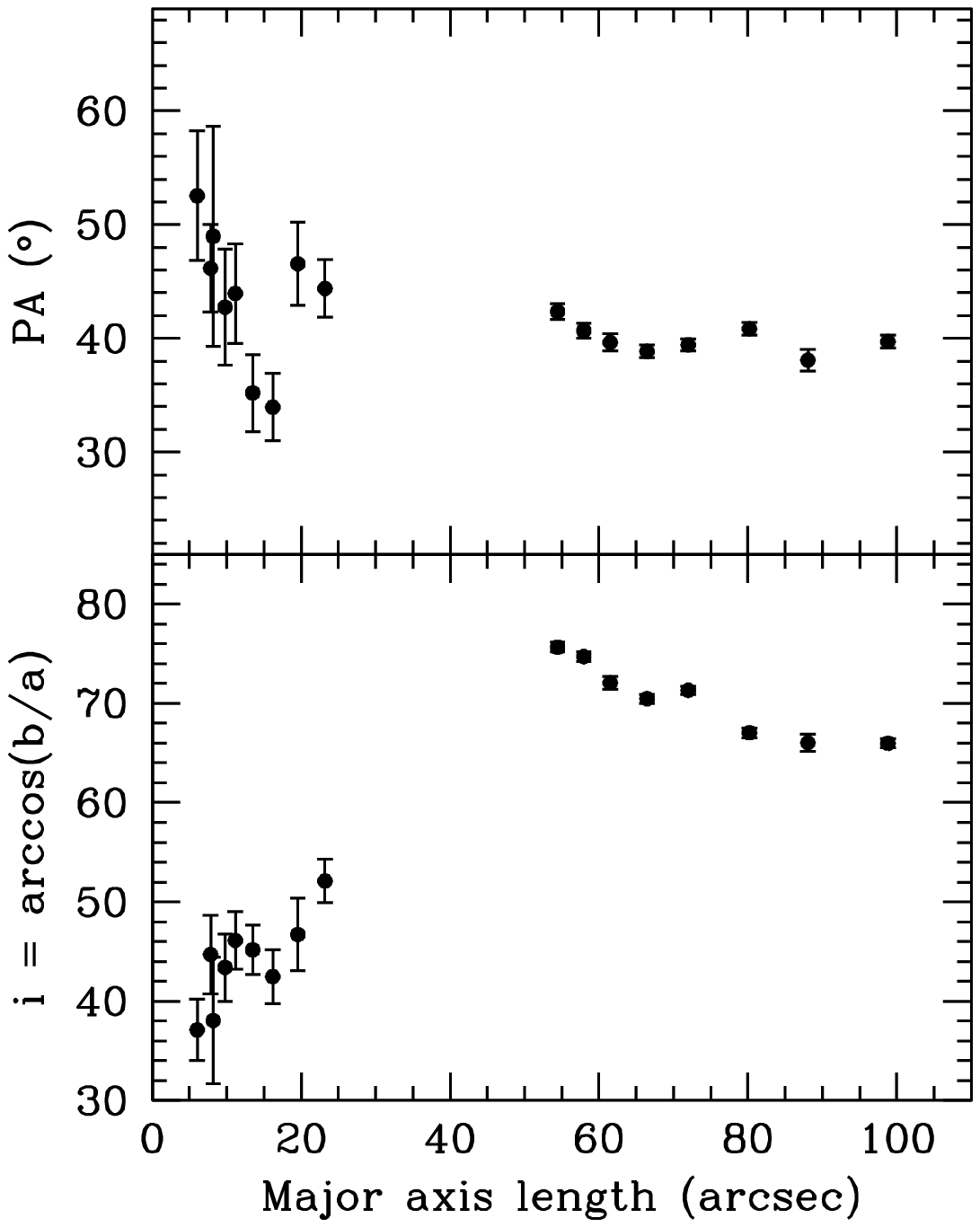}}
\caption[]{The orientation and inclination of ellipses fitted to the
$\kprime$ band image of NGC 3198.  The same fixed central positions have been
adopted for all ellipses.  The distinct signature of the central bulge
(bar) is apparent.
}
\end{figure}

It is without doubt that NGC 3198 has a bulge or central light concentration.
A possible bulge/disc light decomposition is shown in Fig.\ 7.
Here, it is assumed that the 
mass surface density is exactly proportional to the observed intensity level,
and that the light and mass distribution are axisymmetric.
For that case the bulge/disc decomposition illustrated in Fig. 7
is essentially a decomposition by eye. Here, it is further assumed that the
stellar disc has a central hole with a radius corresponding to that of
the light depression and that
the bulge extends slightly beyond this radius.  

\begin{figure}
\resizebox{\hsize}{!}{\includegraphics{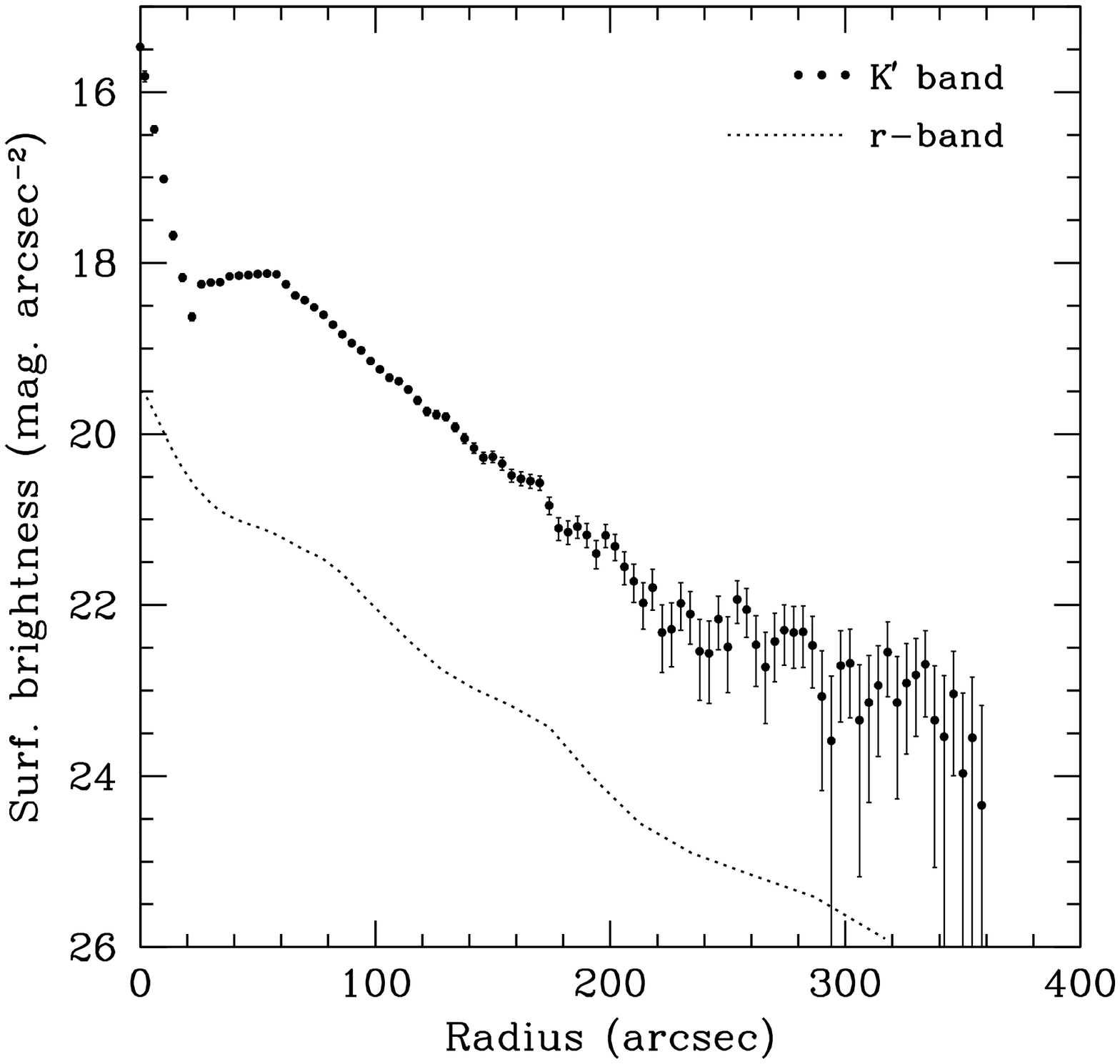}}
\caption[]{
Radial light profile in the $\kprime$ band compared with that in
the r-band (Kent 1987). The photometry of the disc is similar
for both bands, however, in the near infrared the bulge is
considerably brighter than in the optical.
}
\end{figure}

The light depression might well be caused by the presence
of a central bar.  The influence of a bar on the radial velocity field
of the gas is suggested in high-resolution H$\alpha$ images of the 
galaxy, where characteristic distortions from circular motion are 
evident (Corradi et al. 1991).  The light depression
would then be due to a real deficiency of matter near the L4 and L5 
Lagrangian points along the minor axis of the bar (Bosma 1978).  In that
case, the bar would be oriented nearly parallel to the line-of-sight
and would not be photometrically conspicuous.  Moreover, the bar would
affect the derived rotation curve in the inner regions, or, at least, 
the interpretation of the rotation curve as a tracer of the radial force
distribution.  A bar aligned with the minor axis of the galaxy image would
have the effect of increasing the  apparent rotation velocities in the
inner region (Teuben \& Sanders 1985); however, this would be significant
only within the inner 30 arc seconds ($\approx 2$ kpc) 
and would have little influence upon
the overall shape of the derived 21 cm line rotation curve.

\begin{figure}
\resizebox{\hsize}{!}{\includegraphics{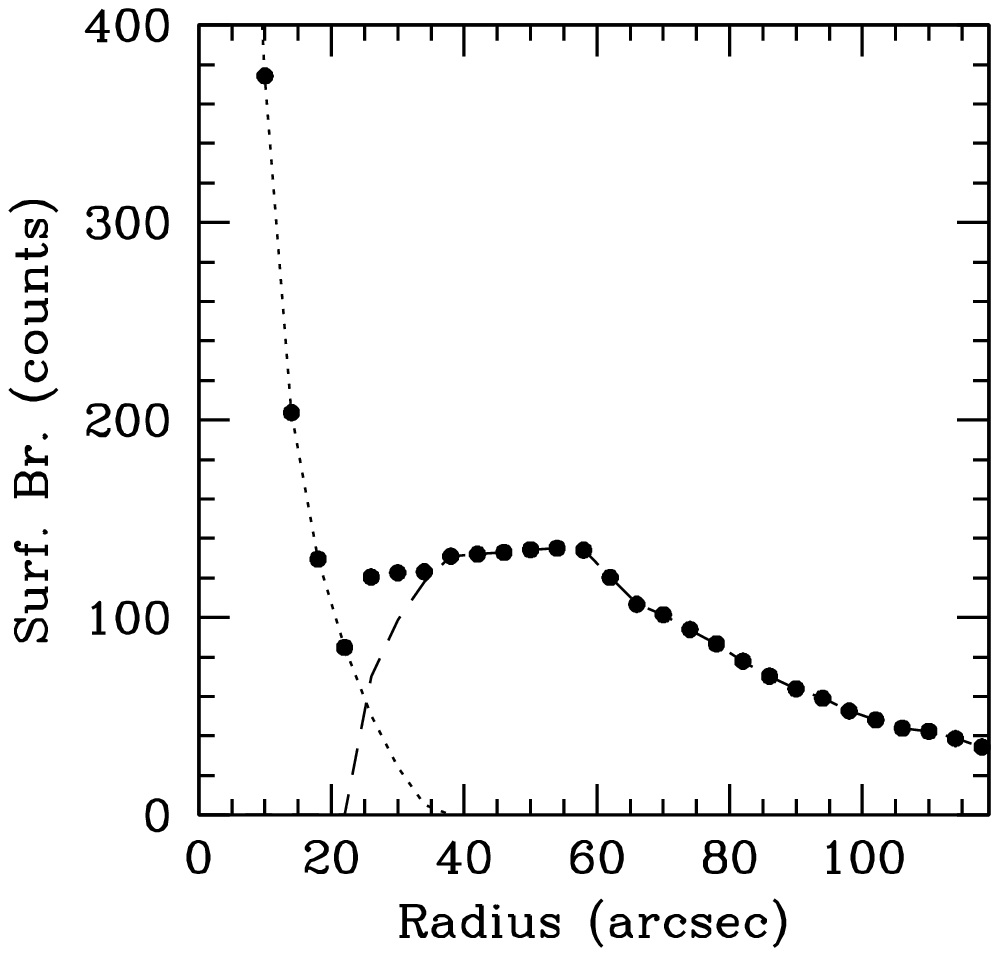}}
\caption[]{
Radial light profile in the central regions and a possible
decomposition into a bulge (dotted line) and a disc (dashed line).
}
\end{figure}

Keeping this caveat in mind, we proceed using the decomposition 
depicted in Fig.\ 7:  assuming a
spherical bulge and disc with observed ellipticity, the total
$\kprime$ luminosity of 3.4~10$^{10}$ $L_{\sun}^{\kprime}$, is
divided into 8.23~10$^9$ and 29.09~10$^9$ $L_{\sun}^{\kprime}$
for the bulge and disc respectively.

Because the scale-length of the disc in $\kprime$ is nearly equal
to the disc scale-length in the optical, it is not to be expected
that the MOND fit will be much different from that for the
r-band photometry. This is the case, as can be seen in Fig. 8 
where the MOND rotation
curve again has been determined at the Cepheid-based distance of
13.8 Mpc. Here, except for a spike in the central regions which is
due to the bulge, the predicted rotation curve is essentially
the same as derived from the r-band photometry;  that is to say, the
conclusions are unchanged by the near-infrared results.

\begin{figure}
\resizebox{\hsize}{!}{\includegraphics{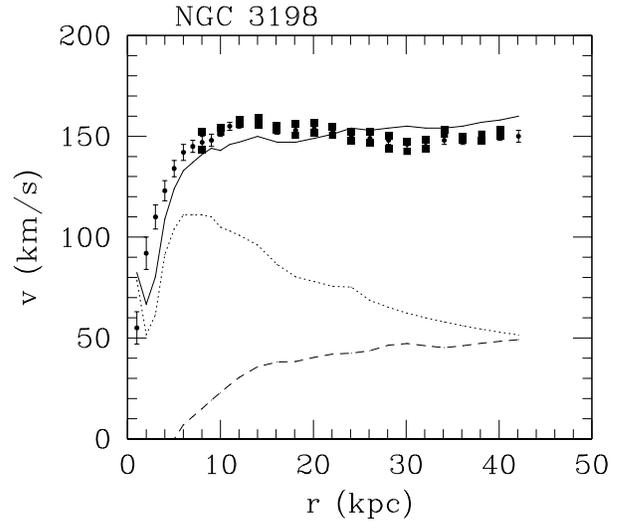}}
\caption[]{
The MOND rotation curve of NGC 3198 where the surface density distribution
of the stellar component is taken to be traced by the $\kprime$-band photometry using
the decomposition shown in Fig.\ 7.  
The distance is Cepheid-based distance of 13.8 Mpc.
The filled squares show the rotation curves derived for 
the two sides of the galaxy
independently; this gives a better estimate of the errors.
}
\end{figure}

\section{NGC 2841}

In Fig.\ 9 we show the MOND rotation curve for NGC 2841 compared to
the observed curve at various assumed distances:  15.6 Mpc, which
is the 1$\sigma$ upper limit on the Cepheid-based distance,
17 Mpc which is 20\% larger than the Cepheid-based distance,
and 23 Mpc which is the MOND-preferred distance.  As in Fig.\ 2 the
M/L values for the disc and bulge are also given in the figure.

\begin{figure}
\resizebox{\hsize}{!}{\includegraphics{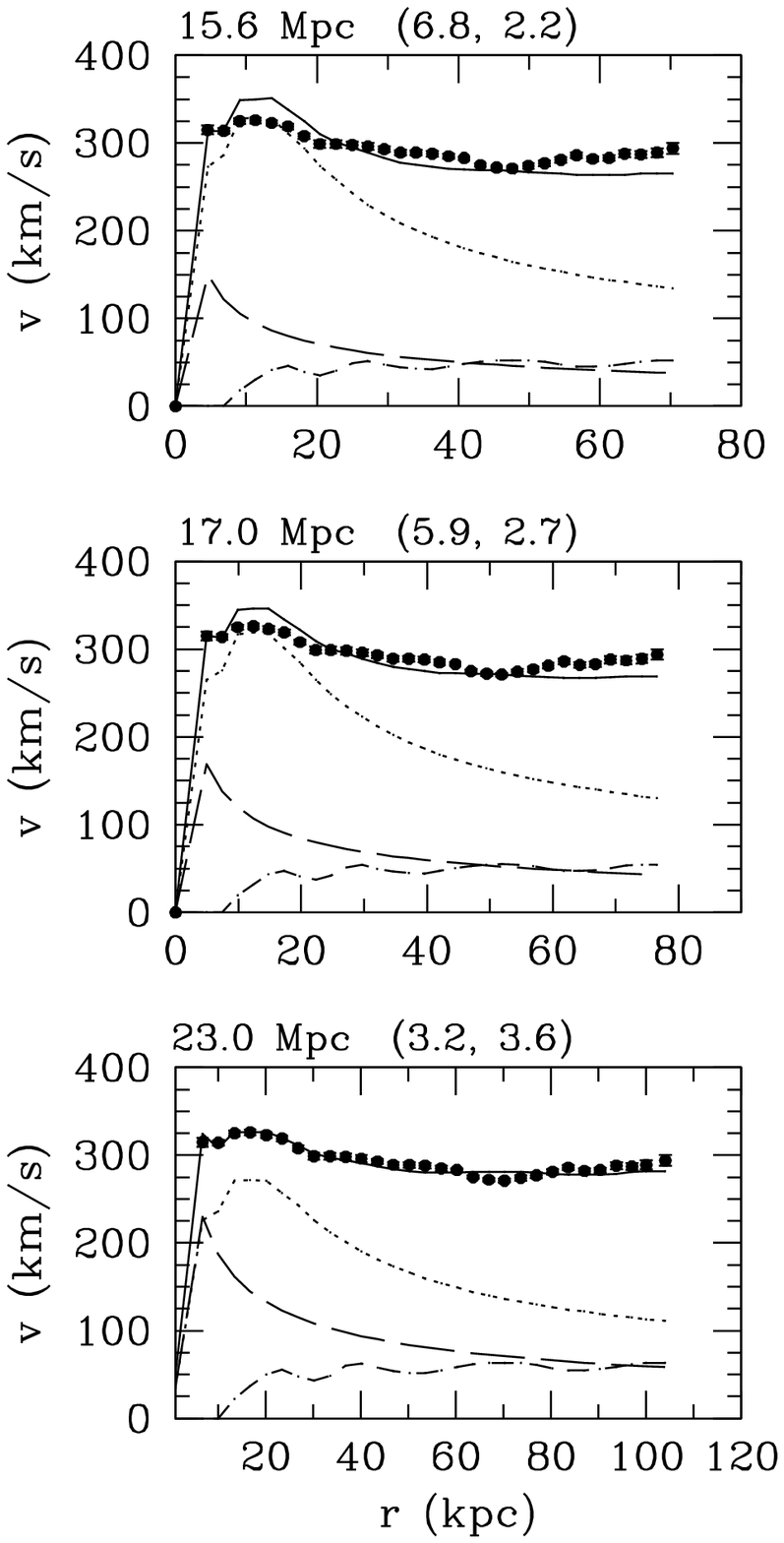}}
\caption[]{
MOND rotation curves for NGC 2841 at various distances ranging from
the one-sigma upper limit on the Cepheid-based distance (15.6 Mpc) to
the Tully-Fisher and SNIa distance (23 Mpc).  The ${\rm M/L_B}$
values for the disc and bulge respectively are given in parenthesis.
}
\end{figure}

The rotation curve, as a tracer of the radial force distribution in 
this galaxy, is actually not as well-determined
as that of NGC 3198.  There is a significant warp in the outer regions
which must be modelled by the tilted-ring technique, and this adds
uncertainty to the derived rotation curve (see comments by Bosma 2002).  
None-the-less, it is
clear that, while the Cepheid distance goes in the right direction
(it is significantly larger than the Hubble based distance), it is not
enough to bring the MOND-predicted rotation curve into agreement with
the observed curve.  Moreover, not only does the form of the predicted
curve differ systematically from that observed, but it is clear that 
the M/L value for the disc is un-naturally large (6.8 $M_\odot/L_\odot$)--
larger than that required for the bulge (2.2).

Both of these problems are relieved somewhat if the distance is taken to be 
20\% larger than the Cepheid-based determination-- at 17 Mpc.  There are
still systematic deviations in the form of the rotation curve, but these
become large (in the outer regions), only where the gas layer of the
galaxy is observed to be significantly warped.  We may take this as
an lower limit on the distance which would be compatible with MOND, although
the disc M/L does remain uncomfortably large (5.9 $M_\odot/L_\odot$) and
that of the bulge rather small (2.7).

Leaving distance as a free parameter in the fit yields a MOND-preferred
distance of 23 Mpc, and here we see that the rotation curve fit is perfect
with very reasonable implied M/L values for the disc and bulge.
This is entirely consistent with the distance implied by the
Cepheid-calibrated Tully-Fisher relation as is also shown in Fig.\ 3.
Taking the galaxy to be at the Cepheid distance of 14.1 Mpc, we see
that the galaxy lies about one entire magnitude below the mean line
of the TF relation.  The distance implied by I-band TF relation 
is 24 Mpc.

There has been a recent supernova in NGC 2841 (SN 1999by), which is
type Ia, i.e., the fundamental extragalactic ``standard candle". 
However, if the galaxy is at the Cepheid distance of 14.1 Mpc, 
SN 1999by is one of the least luminous supernovae Ia ever observed,
with a peak absolute magnitude of M$_B$ = -17.15 $\pm$ 0.23.  Based upon
an estimate of the decline-rate parameter ($\Delta$m$_{15}\approx 1.9$) 
Garnivich et al. (2001) argue that this supernova is a peculiar low 
luminosity event, and they use this event and several others to 
recalibrate the Phillips relation (Phillips et al. 1999) between decline
rate and peak luminosity.  However, if we take the Phillips relation
at face value then the peak luminosity of this object would
be M$_B$ = -18.3, which would imply that the distance to the galaxy
would be 23.5 Mpc.  It is interesting that an earlier SN event in
NGC 2841, SN 1957A, would be, if the galaxy is at the Cepheid distance,
the faintest supernova type Ia ever observed (M$_B$ = -16.4).  It is
curious that this galaxy only seems to provide sub-luminous
supernovae.

The deviation of the galaxy from the TF relation and the abnormally
low peak powers of supernovae, suggest that the the Cepheid distance
to this object may be substantially too low.  It has been argued
that the Cepheid method may be adversely affected by blending:  the true
apparent brightness of Cepheids is enhanced by blending with the light of
nearby stars.  
This would lead to an underestimate of distances based
upon the period-luminosity relation, and would affect, in particular,
the more distant objects (see Paczy\'nski \& Pindor 2001
for a discussion of these points).  All we can conclude, at the moment, is that
the MOND-preferred distance to NGC 2841 remains significantly larger than
the present Cepheid-based distance.

\section{Conclusions}

The main conclusions of this paper can be summarized as follows:

\begin{enumerate}
\item
For the galaxies NGC 2403 and NGC 7331, 
MOND rotation curves agree acceptably with the observed
curves when these galaxies are taken at the Cepheid-based distance.
\item
For NGC 3198 at the Cepheid-based distance of 13.8 Mpc, the MOND
curve shows small ($<10$ \kms) but significant systematic deviations
from the observed curve.
\item
If the distance to NGC 3198 is taken to be 12.5 Mpc, or 10\% less
than the Cepheid-based distance, the MOND curve is an acceptable
representation of the observed curve.  This lower distance 
is probably within the uncertainties of the Cepheid method.
\item
These conclusions are unaltered by utilizing recent 
near-infrared photometry of NGC 3198 which does show evidence for
a small central bulge and bar component.
\item
For NGC 2841, the rotation curve predicted by MOND when the galaxy
is taken to be at the upper limit on the Cepheid-based distance (15.6 Mpc)
remains inconsistent with the observed curve, with systematic deviations
of more than 30 \kms.  
\item
The smallest distance for which the MOND curve is compatible with the
observed curve (given the uncertainties involved in the tilted ring
technique for modelling warps), is 17 Mpc or 20\% larger than the
Cepheid-based distance.  The preferred MOND distance is 23 Mpc.
\item 
The TF distance to NGC 2841, based upon the Cepheid-re-calibrated TF 
relation is 24 Mpc.  If the distance to this galaxy is really 14.1 Mpc,
then it would deviate from the mean I-band TF relation by more than 
1 magnitude.  
\item 
NGC 2841 has been the host of a type Ia supernova, 1999by.  If this
galaxy is at the Cepheid-based distance, this would be one of the
least intrinsically luminous supernovae ever observed.  Calibrating
the peak luminosity by the Phillips relation, the SN-based distance
is 23.5 Mpc.
\end{enumerate}

It is clear that NGC 2841 remains a critical case for MOND.  
The discrepancy between Cepheid-based distance and both the TF and
SNIa based distances to NGC 2841 suggests that there may be a problem with
the derived Cepheid-based distance.  

In general, it is 
evident that accurate distance determinations to nearby galaxies are
extremely relevant to the question of the viability of MOND. MOND,
as a modification of Newtonian dynamics attached to an acceleration
scale, is far more fragile than
the dark matter hypothesis in this regard.  It would
be useful to obtain more Cepheid-based distances to the sample
of galaxies with well-observed rotation curves.  Particularly useful would
be a Cepheid distance estimate to the Ursa Major cluster 
as many of these
galaxies have well-measured rotation curves and near-infrared photometry.

\begin{acknowledgements}
We are grateful to M. Milgrom for a critical reading of the manuscript
and we thank the referee A. Bosma, for giving useful comments which
have improved the paper. 
\end{acknowledgements}


\begin{thebibliography}{}

\bibitem [Begeman 1987]{bg87} Begeman, K.G. 1987, PhD Dissertation,
    Univ. of Groningen
\bibitem [Begeman et al. 1991]{BBS} Begeman, K.G.,
    Broeils, A.H. \& Sanders, R.H. 1991, MNRAS, 249, 523 (BBS)
\bibitem [Bosma 1978]{Bos78} Bosma, A. 1978, PhD Dissertation, Univ.
    of Groningen
\bibitem [Bosma 2002]{Bos01} Bosma, A. 2002, in {\it The Dynamics,
    Structure and History of Galaxies}, ASP Conf. Series 273, Eds.
    da Costa, G. \& Jerjen, H., p. 223
\bibitem [Corradi et al. 1991]{coeal91} Corradi, R.L.M., Boulesteix, J.,
    Bosma, A., et al. 1991, A\&A, 244, 27
\bibitem [Freedman \& Madore 1988]{fm88} Freedman, W.L. \& Madore, B.F.
    1988, ApJ, 326, 691
\bibitem [Freedman et al. 2001]{freal01} Freedman, W.L.,
Madore, B.F., Gibson, B.K., et al. 2001,
    ApJ, 553, 47
\bibitem [Garnavich et al. 2001]{geal01} Garnavich, P.M.,
Bonanos, A.Z., Jha, S., et al. 2001,
    astro-ph/0105490
\bibitem [Hughes et al. 1998]{hueal98} Hughes, S.M.G.,
Han, M., Hoessel, J., et al. 1998,
    ApJ, 501, 32
\bibitem [Hunter et al. 1986]{hueal86} Hunter, D.A., Rubin, V.C. 
\& Gallagher, J.S. 1986, AJ, 91, 1086
\bibitem [Kelson et al. 1999]{keal99} Kelson, D.D.,
Illingworth, G.D., Saha, A., et al. 1999,
    ApJ, 514, 614
\bibitem [Kent 1987]{kt87} Kent, S.M. 1987, AJ, 93, 106
\bibitem [McGaugh \& de Blok 1998]{mcdb98} McGaugh, S.S. \& de Blok, W.J.G.
    1998, ApJ, 499, 66
\bibitem [Macri et al. 2001]{maceal01} Macri, L.M.,
Calzetti, D., Freedman, W.L., et al. 2001,
    ApJ, 549, 721
\bibitem [Maoz et al. 1999]{maoeal99} Maoz, E.,
Newman J.A., Ferrarese, L., et al. 1999, Nature,
    401, 351
\bibitem [Milgrom 1983]{m83} Milgrom, M. 1983, ApJ, 270, 365
\bibitem [Paczy\'nski \& Pindor 2000]{pp01} Paczynski, B. \& Pindor, B.
    2000, ApJ, 533, L103
\bibitem [Phillips et al. 1999]{phill99} Phillips, M.M., Lira, P.,
    Suntzeff, N.B., et al. 1999,
    ApJ, 118, 1766
\bibitem [Rothberg et al. 2000]{r00} Rothberg, B., Saunders, W.,
    Tully, R.B. \& Witchalls, P.L. 2000, ApJ, 533, 781
\bibitem [Sakai et al. 2000]{sak00} Sakai, S.,
Mould, J.R., Hughes, S.M.G., et al. 2000, ApJ,
    529, 698
\bibitem [Sanders 1996]{rhs96} Sanders, R.H. 1996, ApJ, 473, 117 
\bibitem [Sanders \& Verheijen 1998] {sv98} Sanders, R.H. \& Verheijen    
    M.A.W. 1998 ApJ, 503, 97
\bibitem [Teuben \& Sanders 1985]{ts85} Teuben, P.J. \& Sanders, R.H.
    1985, MNRAS, 212, 257
\bibitem [Trimble 1987]{tr87} Trimble, V. 1987,
    ARAA, 25: 425
\bibitem [Tully \& Fisher 1977]{tf77} Tully, R.B. \& Fisher, J.R.
    1977, A\&A, 54, 661
\bibitem [Tully \& Pierce 2000]{tp00} Tully, R.B. \& Pierce, M.J. 2000,
    ApJ, 533, 744
\bibitem [van Albada et al. 1985]{vaeal85} van Albada, T.S., Bahcall, J.N.,
    Begeman, K. \& Sancisi, R. 1985, ApJ, 295, 305
\bibitem [van der Kruit \& Searle 1981]{vdks81} van der Kruit, P.C.
    \& Searle, L. 1981, A\&A, 95, 105
\end{thebibliography}
\end{document}